\input{aipcheck}

\documentclass[
    ,final            
  ]
  {aipproc}

\layoutstyle{8x11double}

\begin{document}

\title{Quantitative Description of Strong-Coupling of Quantum Dots
  in Microcavities}


\author{F.P. Laussy}{
  address={School of Physics and Astronomy, University of Southampton, United Kingdom}
}

\author{E. del Valle}{
  address={Universidad Aut\'onoma de Madrid, Facultad de Ciencias C-V, Cantoblanco, Madrid, Spain}
}

\author{C. Tejedor}{
  address={Universidad Aut\'onoma de Madrid, Facultad de Ciencias C-V, Cantoblanco, Madrid, Spain}
}

\begin{abstract}
  We have recently developed a self-consistent theory of
  Strong-Coupling in the presence of an incoherent
  pumping~[arXiv:0807.3194] and shown how it could reproduce
  quantitatively the experimental data~[PRL \textbf{101}, 083601 (2008)].
  Here, we summarize our main results, provide the detailed analysis
  of the fitting of the experiment and discuss how the field should
  now evolve beyond merely qualitative expectations, that could well
  be erroneous even when they seem to be firmly established.
\end{abstract}

\maketitle


\begin{figure}[h!]
  \centering
  \includegraphics[width=1.68\linewidth]{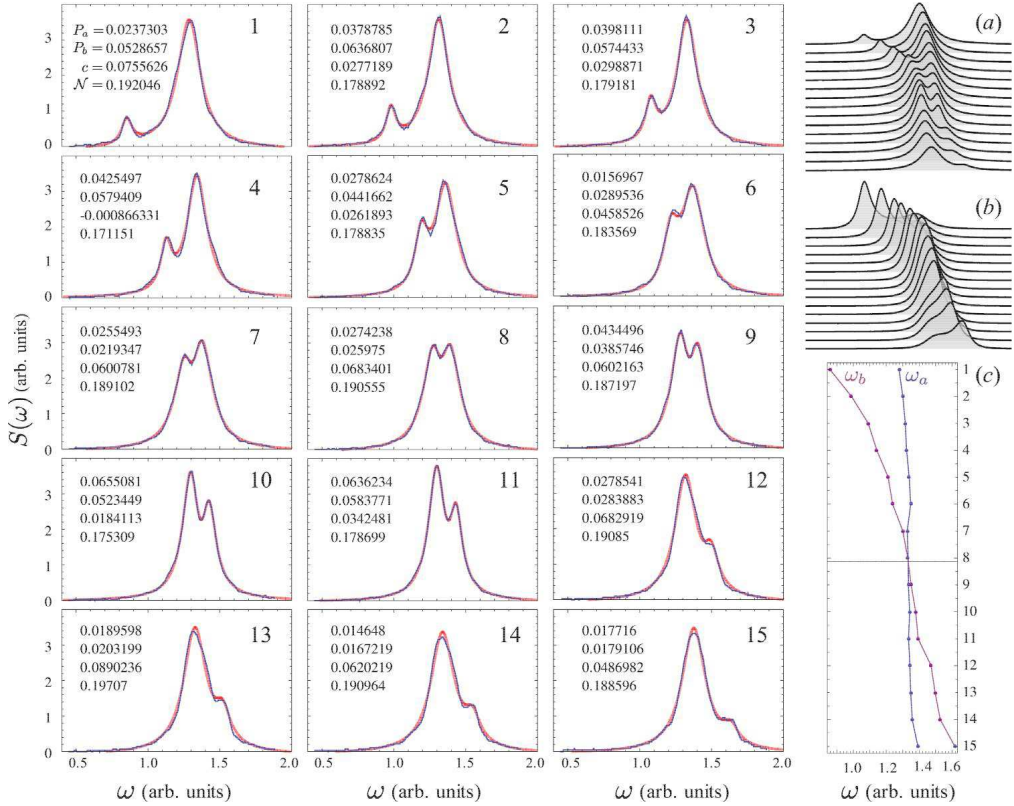}
  \caption{Theoretical fit (in semi-transparent red) of the data by
    Reithmaier \emph{et al.} digitized from Ref.~\cite{reithmaier04a}
    (blue). The model~\cite{qdsc_laussy08a} provides
    $S(\omega)=\frac{1}{2\pi}\mathrm{Re}\sum_{\sigma=-1,1}\frac{1+\sigma\mathcal{C}}{\Gamma_++i[\sigma
      R+\omega-(\omega_a+\omega_b)/2]}$ with
    $\Gamma_\pm=({\gamma_a-P_a\pm(\gamma_b-P_b)})/{4}$,
    $R=\sqrt{g^2-(\Gamma_-+i\Delta/2)^2}$,
    $D={\frac{g}2(\gamma_aP_b-\gamma_bP_a)(i\Gamma_+-{\Delta}/2)}/
    {g^2\Gamma_+(P_a+P_b)+P_a(\gamma_b-P_b)(\Gamma_+^2+({\Delta}/{2})^2)}$
    and $\mathcal{C}={\Gamma_-+i(\Delta/2+gD)}/R$. The data has been
    fitted on rescaled axes for numerical stability by a
    Levenberg--Marquardt method with $\mathcal{N}S(\omega)-c$,
    with~$\mathcal{N}$ and~$c$ to account for the normalization and
    the background. Beside these two necessary parameters regardless
    of the model, each panel only has~$P_{a/b}$ and~$\omega_{a/b}$ (c)
    as fitting parameters. $g$ and $\gamma_{a/b}$ have been optimized
    globally, with best fits for~$g=61\mu$eV, $\gamma_a=220\mu$eV
    and~$\gamma_b=140\mu$eV.  (a) shows the anticrossing from
    curves~1--15 put together. (b) keeps all fitting parameters the
    same but with~$P_a=0$ and vanishing~$P_b$. The dot emission now
    dominates and no anticrossing is observed, although the system is
    still in strong-coupling.}
  \label{fig:ThuJul24210148CEST2008}
\end{figure}

\emph{Strong-Coupling} (SC) is the term consecrated by the quantum
optics community to designate ``quantum coupling'', where coherent
interaction dominate over dissipation.  A vivid picture represents
this regime in terms of a cyclic \emph{exchange} of energy between the
light and matter fields (the so-called ``\emph{Rabi
  oscillation}''). This oscillation is a particular case as can be
seen straightforwardly by considering two conceptual cases where it
does not occur: when the system is in an eigenstate of the SC
hamiltonian, in which case it has no dynamics, and when it has reached
a steady state (SS), where, regardless of which quantum state is
realized, energy is balanced rather than exchanged. The SS is the
relevant case for many experimental configurations with
semiconductors~\cite{reithmaier04a,yoshie04a,peter05a}, where a
continuous incoherent pumping excites the system. An anticrossing is
looked for in the luminescence spectra as a proof of SC, attributed to
the emergence of ``dressed'' states.

To compute a luminescence spectrum of a QD in a semiconductor, one
must consider an innocent-looking but important consequence of
pumping: it enforces a steady state that is a mixture of cavity
photons and excited state of the QD (exciton).  The lineshape of the
cavity luminescence spectrum~$S(\omega)$ depends strongly on whether
the system is \emph{photon-like} or
\emph{exciton-like}~\cite{qdsc_laussy08a}, in the sense of which
particle would be injected in the system in the limit of vanishing
pumpings that keep a fixed ratio as they go to zero.  The steady state
should be computed self-consistently from the interplay of pumping and
decay, rather than assuming that the system is, typically, in the
excited state of the QD. The latter assumption would be reasonable if
there was only one kind of pumping, namely in this case an electronic
pumping. However, experimental evidence seems to imply that an
incoherent cavity pumping accompanies the electronic
pumping~\cite{qdsc_laussy08b}. The most likely reason is that beside
the QD that is strongly coupled to the cavity, there are many other
dots in weak-coupling, that also get excited by the electronic
pumping, and that populate the cavity when they de-excite. This, for
the SC dot, is perceived as an incoherent cavity pumping.

Even in the simplest description where the two modes (light, $a$, and
matter, $b$) are bosonic, the luminescence spectra enjoy a wealth of
subtle characteristics when incoherent and continuous pumpings
$P_{a/b}$ for the cavity/exciton, are introduced. Those are discussed
at length in Ref.~\cite{qdsc_laussy08a}. Here we mention the most
important ones: anticrossing is not systematically observed for a
system in SC, depending on whether the system is more photon-like
($P_a>P_b$), or more exciton-like ($P_b>P_a$). More surprisingly, an
``apparent'' anticrossing can be observed in a system that is in
weak-coupling, due to an interference that carves a hole in the
spectra.  These two facts together disconnect completely an
anticrossing behaviour from the realization of SC. One should not,
therefore, rely on this qualitative effect to ascertain SC. Instead, a
fitting of the data should be attempted, so as to place a given
experiment in a region of SC defining rigorously the dressed states,
rather than from their ability to survive in the luminescence
spectrum.

We have considered one of the pioneering
experiment~\cite{reithmaier04a}, as a test of the model, and found an
excellent agreement with a linear (bosonic) model,
cf.~Fig.~\ref{fig:ThuJul24210148CEST2008}, provided that both cavity
and exciton pumping are taken into account. This is consistent with
the experimental curves that feature a strong cavity emission. In
panel~(b) we show how the lineshapes transform when the cavity pumping
is switched off: the exciton line dominates, and no anticrossing is
observed. The system is nevertheless still in SC. The confrontation of
the theory with the experimental data was done for illustration only,
as the raw experimental data was not available by the time of our
investigation. We have therefore digitized the data, what forbids an
in-depth statistical analysis, since the experimental points are
required rather than the interpolated curves published in
Ref.~\cite{reithmaier04a}, if only to know the numbers of degrees of
freedom. Note that one expects better still results as our procedure
added noise. We found the best agreements near resonance, which might
be due to the exciton that, when it is less-strongly coupled at larger
detunings, may go below the resolution of the detector, resulting in
an apparently broader line. All these limitations can be circumvented
with a careful statistical analysis (and treatment of the data to
reconstruct linewidths below the experimental resolution). This is a
standard procedure of a mature field to validate a theoretical model
over another by statistical analysis of the experiment. Nullifying
hypotheses such as: ``\emph{a Fermi (two-level) system accounts for
  the observed data better than a Boson model}'' are important to
assess the achievements made in terms of quantum emitters with these
systems. This would also provide a meaningful and quantitative
comparison between the various implementations (micropillars,
microdisks and photonic crystals). Lacking the full experimental data,
we have merely been unable to provide a confidence interval to our
most-likelihoods estimators. Doing so, progress will be meaningfully
quantified, and claims---rather than ranging between likely and
convincing---will become unambiguously proven (within their interval
of confidence).


\end{document}